\documentclass[12pt]{article}
\usepackage[top=1in, bottom=1in, left=1.in, right=1.in]{geometry}
\usepackage[english]{babel}
\usepackage{atbegshi,cite}
\usepackage{amsmath,amssymb,amsbsy,amstext, amsthm, simplewick}
\usepackage{hyperref}
\usepackage{graphicx}
\usepackage{amsfonts}
\usepackage[small]{caption}
\usepackage{upgreek}
\usepackage[titletoc]{appendix}
\usepackage[usenames,dvipsnames,table]{xcolor}
\usepackage{setspace}
\usepackage{color}

\newcommand{\newc}{\newcommand}
\newc{\fpi}{f_{\pi}}
\newc{\etap}{\eta^{\prime}}
\newc{\llll}{\langle\lambda\lambda\rangle}
\newc{\FFd}{F^a\tilde F^a}
\newc{\qbar}{{\overline q}}
\newc{\TR}{{\rm Tr}}
\newc{\Kahler}{K\"ahler }
\newc{\Zbb}{{\mathbb Z}}
\newc{\Rt}{{\mathbb R}^3}
\newc{\Rf}{{\mathbb R}^4}
\newc{\So}{{\mathbb S}^1}
\newc{\zt}{{\mathbb Z}_2}
\newc{\RtSo}{{\mathbb R}^3\times{\mathbb S}^1}
\newc{\scriminus}{{\cal I}^-}
\newc{\scriplus}{{\cal I}^+}
\newc{\mpl}{M_p}
\newc{\Ricci}{\mathcal{R}}
\newc{\bv}{\phi}
\newc{\hatr}{{\hat r}}
\newc{\calK}{K}
\newc{\calUi}{{\cal U}^{-1}}
\newc{\calE}{{\cal E}}
\newc{\calH}{{\cal H}}
\newc{\calL}{{\cal L}}
\newc{\calN}{{\cal N}}
\newc{\calM}{{\cal M}}
\newc{\calI}{{\cal I}}
\newc{\calG}{{\cal G}}
\newc{\calO}{{\cal O}}
\newc{\calU}{{\cal U}}
\newc{\calUp}{{\cal U}^\prime}
\newc{\calUd}{{\cal U}^\dagger}
\newc{\calUpd}{{\cal U}^{\prime \dagger}}
\newc{\calX}{{\cal X}}
\newc{\calXp}{{\cal X}^\prime}
\newc{\calXd}{{\cal X}^\dagger}
\newc{\calXpd}{{\cal X}^{\prime \dagger}}
\newc{\calQ}{{\cal Q}}
\newc{\calOb}{{\cal O}^\dagger}
\newc{\hphi}{{\hat\phi}}
\newc{\llangle}{\langle\langle}
\newc{\rrangle}{\rangle\rangle}
\newc{\?}{\overset{?}{=} }

\theoremstyle{plain}
\theoremstyle{plain} 
\theoremstyle{plain} 
\theoremstyle{plain}
\theoremstyle{plain}
\theoremstyle{plain}

\renewcommand{\title}[1]{{\Large\bf\flushleft{#1}}\vspace*{3ex}\\}
\renewcommand{\author}[2]{{\noindent\hspace*{2.5em}\large#1}
                     \footnote{Electronic mail: $\mathtt{#2}$}\\}

\begin{document}

\begin{titlepage}

\vskip 2.2cm

\begin{center}

{\large \bf   Generalized Entanglement Capacity of de Sitter Space}
\vskip 1.4cm

{Tom Banks$^{(a),}$\footnote{tibanks@ucsc.edu} and Patrick Draper$^{(b),}$\footnote{pdraper@illinois.edu}}
\\
\vskip 1cm
{ $^{(a)}$NHETC and Department of Physics \&\\Rutgers University, Piscataway, NJ 08854-8019}\\
{ $^{(b)}$Illinois Center for Advanced Studies of the Universe \&\\Department of Physics, University of Illinois, Urbana, IL 61801}\\
\vspace{0.3cm}
\vskip 4pt

\vskip 1.5cm
\begin{abstract}
Near horizons, quantum fields of low spin exhibit densities of states that behave asymptotically like 1+1 dimensional conformal field theories. In effective field theory, imposing some short-distance cutoff, one can compute thermodynamic quantities associated with the horizon, and the leading cutoff sensitivity of the heat capacity is found to equal to the leading cutoff sensitivity of the entropy. One can also compute contributions to the thermodynamic quantities from the gravitational path integral. For the cosmological horizon of the static patch of de Sitter space, a natural conjecture for the relevant heat capacity is shown to equal the Bekenstein-Hawking entropy. These observations allow us to extend the well-known notion of the generalized entropy to a generalized heat capacity for the static patch of dS. The finiteness of the entropy and the nonvanishing of the generalized heat capacity suggests it is useful to think about dS as a state in a finite dimensional quantum gravity model that is not maximally uncertain.
\end{abstract}

\end{center}

\vskip 1.0 cm

\end{titlepage}
\setcounter{footnote}{0} 
\setcounter{page}{1}
\setcounter{section}{0} \setcounter{subsection}{0}
\setcounter{subsubsection}{0}
\setcounter{figure}{0}

\section{Introduction}

The quantum theory of de Sitter (dS) space remains obscure. We do not know the nature or even the quantity of the microscopic degrees of freedom, or their interactions.

In~\cite{tbdS,wfdS} it was proposed that the Gibbons-Hawking entropy of dS space could be interpreted as the logarithm of the dimension of a finite-dimensional quantum system describing a static patch of the spacetime.  This hypothesis
implies that empty dS space corresponds to the maximally uncertain density matrix of the quantum system.  Within semiclassical physics, it is supported by the Schwarzschild-de Sitter (SdS) entropy formula, which shows that objects localized near a geodesic reduce the entropy, and by the fact that generic localized excitations melt into the near-horizon region of any given static patch in a time of order $R_{dS}$. Furthermore, attempts to introduce higher-entropy initial states in the remote past of the global dS manifold lead to singular space-times\footnote{This was originally a conjecture made by TB and was later proven in~\cite{Bousso:2005yd}.} rather than future asymptotically-dS space.

The ``maximum entropy" proposal is simple and compelling, and it has received support of a kind from recent progress in the theory of operator algebras. It would be interesting to establish whether it is strictly true, or only holds in some approximation. A natural quantity to examine is the second cumulant of the modular Hamiltonian $K$, where $\rho=e^{-K}$ is the normalized density matrix of the microscopic state corresponding to empty dS.  The entropy satisfies $S = \langle K\rangle$. For a state of maximum entropy, $\rho = \mathbf{I}/\dim H$, and the second cumulant (also referred to as the ``capacity of entanglement"~\cite{DeBoer:2018kvc,Nakaguchi:2016zqi,Nakagawa:2017wis,Nandy:2021hmk}) vanishes, $\langle (K-\langle K\rangle)^2\rangle=0$. 
Thus the 
magnitude of the modular fluctuation can discriminate the maximum entropy state from other possibilities.

This note is a follow-up to previous work~\cite{Banks:2022irh}, in which it was argued that the semiclassical state of empty de Sitter space need not correspond to a microscopic maximum entropy state. Random projections also reproduce the entropy deficit properties of the Schwarzschild-de Sitter (SdS) geometry, for a general class of density matrices, not just the maximally uncertain one. Furthermore there exist density matrices with the properties that $ \langle K\rangle = \log\dim H +\calO(1)$ and $\langle (K-\langle K\rangle)^2\rangle\sim  \langle K\rangle$.

Here we consider two other probes of the  modular fluctuation in the dS static patch. The first is the behavior of effective quantum field theory (EFT) on the static patch background (Sec.~\ref{eft}). EFTs are presumed to have a physical short-distance cutoff. Without knowing the UV completion, one can estimate the sensitivity of quantities like the entropy and capacity of entanglement to the physical UV scales by putting general, unphysical cutoffs on the EFT. 

The second approach we consider is the semi-classical Euclidean functional integral for gravity, where we make a conjecture in Sec.~\ref{Gravity}  about how one should identify and vary the replica index. A modular fluctuation analog to the classical Bekenstein-Hawking entropy can be computed from a derivative with respect to the replica index.  This kind of semi-classical replica trick has been successful in a wide variety of contexts in the AdS/CFT correspondence. Our proposal remains consistent with the idea that states with localized matter in the bulk correspond to constrained states, essentially because thermal states may either be viewed as minimizing the free energy at fixed temperature, or maximizing the entropy with a constraint on the expectation value of the energy.

Both EFT and semiclassical gravity are consistent with a proportionality of the form $\langle (K-\langle K\rangle)^2\rangle=a \langle K\rangle$, with coefficient $a$ of order one (and in fact equal to one in the cases considered here.)  It is straightforward to explain why the SdS argument described above did not ``see" the large modular fluctuation, which we also discuss in  Sec.~\ref{eft}.  We may define a ``generalized capacity of entanglement" $C_{gen}$ as the sum of the EFT and leading-order semiclassical gravity contributions, and in cases where the value of $a$ is the same in both terms, $C_{gen}$ is proportional to the generalized entropy.

In Sec.~\ref{crossed} we discuss the question in the context of operator algebras and the kinds of modifications to the crossed product construction needed to produce a large nonzero $\langle (K-\langle K\rangle)^2\rangle$.  In Sec.~\ref{disc} we conclude with a discussion of the relation between the modular fluctuation and other ideas in static patch holography.

\section{Effective Field Theory}
\label{eft}

Effective quantum field theory on a static patch of rigid dS, or any spacetime with a horizon, contains a density of near-horizon states that is sensitive to the ultraviolet cutoff. For free theories the sensitivity can be easily calculated with a brick wall regulator~\cite{thooft}, or covariant regulators like Pauli-Villars~\cite{Demers:1995dq} or character expansions~\cite{Anninos:2020hfj}. 
For example, for $N$ free bosons in $D$ spacetime dimensions on a spacetime $ds^2=-f dt^2+f^{-1}dr^2 +r^2d \Omega_{D-2}$ one finds a density of near-horizon single-particle modes of the form
\begin{align}
g(\omega) = N\left(\frac{r_H}{\rho_c}\right)^{D-2}X(r_H T_H)+\dots
\end{align}
Here  $r_H$ is a horizon radius, $f(r_H)=0$; $T_H$ is the surface gravity temperature with respect to the timelike Killing vector $\partial_t$, $T_H = |f'(r_H)|/4\pi$; $X$ is a function;  $\rho_c$ is a short-distance cutoff; and $\dots$ denotes terms that are subleading for small $\rho_c$. The key point is that the single-particle density of near-horizon states is independent of energy $\omega$. This sector of the theory behaves like a 1+1-dimensional  CFT of $N\left(\frac{r_H}{\rho_c}\right)^{D-2}$ free bosons. The factor $\left(\frac{r_H}{\rho_c}\right)^{D-2}$ counts angular modes. In the Hartle-Hawking (HH) or Bunch-Davies (BD) state, these modes are thermally excited at a temperature $T_H$ and make a large contribution to the thermal entropy of order $T_H g(T_H)\sim A/\rho_c^{D-2}$. However, they are also strongly coupled to gravity, in the sense that one cannot excite very many of them away from the HH/BD state at once before gravitational backreaction becomes large~\cite{Banks:2024imv}. One is left with the unsatisfactory conclusion that there is a large collection of near-horizon modes, some but not all of which are perhaps well modeled by  effective quantum field theory on a rigid background. 

In any case, we cannot really say that just because low-entropy bulk excitations lower the cosmological horizon area, the state describing empty de Sitter in the relevant microscopic theory of quantum gravity is of maximum entropy. The near-horizon degrees of freedom do not have to be in this state.

Suppose, for example, that we could neglect the backreaction argument of~\cite{Banks:2024imv} and take effective QFT seriously up to a cutoff scale of order the Planck length, $\rho_c\sim \ell_p$. At inverse temperature $\beta$, because the density of states is independent of energy, the near-horizon states contribute to the partition function
\begin{align}
\Delta(\beta F) \propto \int \log\left(1-e^{-\beta \omega}\right) d\omega
\end{align}
which on dimensional grounds scales as $1/\beta$. 
($\beta$ must be kept independent of $T_H$ in the density of states until we are done taking $\beta$ derivatives.)
So the thermal entropy
\begin{align}
S = \beta^2 \partial F/\partial\beta 
\end{align}
and the heat capacity
\begin{align}
C = \beta^2 \partial^2(-\beta F)/\partial\beta^2 
\end{align}
are equal to each other and proportional to $A/G_N$.  In a thermal state, 
\begin{align}
C =  \langle (K-\langle K\rangle)^2\rangle
\end{align}
so the large heat capacity indicates that the  state is far from the maximally uncertain density matrix. 

Furthermore, since $\langle K \rangle=S$,  $C=S$ implies 
\begin{align}
\langle (K-\langle K\rangle)^2=a\langle K \rangle\;,\;\;\;\;\;a=1. 
\end{align}
The conclusion that $a=1$ actually  holds whenever the near-horizon physics is approximately that of a 1+1-dimensional CFT with a cutoff,  which is the case for free massless scalars and spin-$1/2$ fields, among others. $C$ and $S$ are both sensitive to the UV cutoff, but the ratio, $a$, is finite. In general the value of $a$ may depend on the details of the QFT.

 One way of understanding the apparent $1 + 1$ dimensional nature of the near horizon density of states is the observation of~\cite{Carlip} and~\cite{solo} that the near horizon geometry of dS space, similarly to non-extremal causal diamonds, has an approximate Virasoro symmetry.  Any regularization procedure that respects this symmetry will appear to have the density of states of a $1 + 1$ CFT.   Of course, this might be misleading since any higher dimensional field theory will produce an infinite number of two dimensional free fields. This of course reflects the fact that the entanglement entropy has a power law UV divergence, with a power that depends on the dimension of space-time.  Our point however is that the same divergence shows up in both the entanglement entropy and the capacity of entanglement, and that the ratio is finite, with $a = 1$.

An alternative regulator for the entanglement entropy of QFT has been proposed by~\cite{CHM}.  This elegant method preserves the entire conformal group of the higher dimensional field theory, but explicitly breaks the near horizon Virasoro symmetry. It converts the UV divergence of the entanglement entropy into the volume divergence of the thermal entropy on a hyperbolic space.  Again the ratio between the entanglement entropy and the capacity of entanglement is finite, but now it depends on the particular higher dimensional CFT one is studying.  Curiously, whenever that CFT has an Einstein-Hilbert dual in the AdS/CFT correspondence, the CHM procedure predicts $a = 1$, confirming the calculation of~\cite{VZ2} for boundary anchored Ryu-Takayanagi diamonds in quantum gravity.

\section{Euclidean Gravity and the Replica Trick}
\label{Gravity}

The contribution to the entropy from near-horizon QFT modes, $S_{QFT}$, has a gravitational analog $A/4G_N$. The leading UV-cutoff sensitivity of $S_{QFT}$ generally coincides with the  perturbative renormalization of $G_N$, so that the sum of the two contributions, at leading order, is just $A/4G_N$ with the renormalized Newton constant. We would like to know if a similar picture holds for the modular fluctuation / heat capacity, for which we need a computation of this quantity in pure Euclidean gravity.

One approach is to use replica methods. Let the microscopic density matrix of empty dS space be $\rho = e^{-K}$, with $\TR(\rho)=1$, and define $\rho_n = e^{-nK}/\TR(e^{-nK})$. Then
 \begin{align}
 S_n \equiv -\TR\left( \rho_n\log\rho_n\right)
 \label{Sndef}
 \end{align}
 The modular fluctuation can be shown to satisfy
 \begin{align}
  \langle (K - \langle K \rangle)^2 \rangle = -S'_1.
  \end{align}
The angle brackets on the left-hand side denote expectation values with respect to $\rho$. 

Most applications of replica methods to quantum gravity begin with a boundary in Euclidean signature. First one constructs $n$ copies of the boundary, and the boundary data, glued together in a chain. Then, for each $n$, one looks for smooth interior solutions to the semiclassical gravitational field equations~\cite{LM}. The solution $\calM(n)$ of the lowest action $\calI(n)$ provides an approximation to the $n$-fold partition function of the microscopic quantum gravity theory, $-\log Z(n)\approx \calI(n)$. This is because the boundary is where a nongravitational holographic dual theory lives, and replicating the boundary geometry $n$ times corresponds to constructing  $Z(n)$ in the dual theory~\cite{LM}.

It is not immediately obvious how to  apply similar techniques to learn about quantum gravity ``in de Sitter space." First, the Euclidean continuation of de Sitter is a sphere, which has no boundary. Second, if a holographic dual description lives anywhere, it probably lives at the horizon of the static patch, and the Euclidean continuation of a null surface is of codimension two. One might attempt to place some boundary data on a ``stretched horizon" a small fixed distance above the horizon. There is a problem with  the most na\"ive application of this idea, however, which will be discussed below.

Nonetheless it is widely believed that there is some microscopic theory of quantum gravity with a density matrix $\rho$, for which various properties of $\rho$ -- like the von Neumann entropy -- are computed in some approximation by the Euclidean de Sitter saddle point of the  gravitational path integral. One would like to know whether other suitable saddle points compute $\rho_n$.  Here we make a conjecture based on the Schwarzschild de Sitter (SdS) geometries.

The SdS  metric in $D$ space-time dimensions is
\begin{equation} ds^2 = - f(r) dt^2 + \frac{dr^2}{f(r)} + r^2 d\Omega_{D-2}^2 , \end{equation}
\begin{equation} f(r) = 1 - \left(\frac{r_s}{r}\right)^{D-3} - \frac{r^2}{L^2} . \end{equation} 
$f(r)$ has two positive zeros, and for $r_s \ll L$ one is approximately at $r_b\approx r_s$ and the other at $r_c\approx L - \frac{r_s^{D-3}}{2 L^{D-4}} $. 
As a consequence, the effect of inserting a localized mass distribution near the geodesic, whether a black hole or a ``star," is to shrink the area of the cosmological horizon. Thus, according to the Bekenstein-Hawking-Gibbons law, the entropy of empty dS space is also reduced.  This was interpreted~\cite{tbdS,wfdS,bfm} as an explanation of the Gibbons-Hawking temperature of dS space as a purely entropic effect, arising from a constraint on the holographic degrees of freedom on the horizon. Semiclassically, the constraint takes the form of a projection onto fields of fixed quasilocal energy~\cite{Draper:2022xzl,Morvan:2022ybp,Draper:2023bhg}. This is one bit of an enormous amount of evidence that most of the quantum states of dS space are localized near the horizon, relatively inaccessible to experimental probes, and have static energies of order $1/L$, as indicated by the redshift in static coordinates.

As described above, the conventional idea of the gravitational replica trick in AdS space is to find a Euclidean manifold, which is a smooth solution of Einstein's equations and has periodicity around a thermal cycle on the boundary that is $n$ times as large as the original manifold on which the boundary field theory lives.    Our proposal for dS space is motivated by the fact that the Euclidean SdS manifold changes the periodicity of the Euclidean static time coordinate as it changes the entropy.  We thus define the replica index as the ratio of cosmological horizon surface gravity temperatures (for $r_s \ll L$),
\begin{align}
 n \equiv  1 + \frac{D - 2}{2}\left(\frac{r_s}{L}\right)^{D-3}.
 \label{nrs}
 \end{align}
   The entropy of the cosmological horizon of SdS is
   \begin{align}
S = S_{dS} \left[1 - \frac{D - 2}{2}\left(\frac{r_s}{L}\right)^{D-3}\right]. 
\label{SdSentropy}
\end{align}
  Thus, with the identification of $r_s/L$ and the replica index in Eq.~(\ref{nrs}), and identifying the SdS entropy in Eq.~(\ref{SdSentropy}) with $S_n$ in Eq.~(\ref{Sndef}), we find
  \begin{align}
  S_n &= S_{dS} [1 - (n - 1)],\nonumber\\
  -S'_1 &= S_{dS}.
  \end{align}
Thus we find $a = 1$, independent of $D$. This extends the semiclassical approach of \cite{VZ2} for RT diamonds in AdS to dS, and is in agreement with the results of \cite{BZ}  (arrived at by other means, extrapolating the ideas of  \cite{Carlip,solo}).   Note that we could have multiplied our definition of $n$ by any function $h(r_s / L)$ such that $h(1) = 1$ and $h^{\prime} (1) = 0 $ without changing the result.   This is in keeping with expectations from AdS/CFT that higher order fluctuations do not necessarily take on a universal form.   

Let us address some technical points.  The Euclidean SdS manifold is singular on the continued black hole horizon.  If we add any sort of matter fields that produce a spherically symmetric ``star" solution, the singularity disappears.  We can deal with it simply by cutting out an artificial boundary around the singularity and imposing Brown-York microcanonical boundary conditions\cite{BY,Miyashita:2021iru,Draper:2022ofa}, so that nothing depends on the details of what is going on near the geodesic, and the on-shell action simply computes the cosmological horizon entropy.  Of course, since we take $r_s \rightarrow 0$ at the end of the calculation, this is not a serious issue in any event.

A more subtle point is the normalization of the Killing vector $\partial_t$ in a uniform, $r_s$ independent way.  The traditional way to do this~\cite{GH} extrapolates the static coordinates to an asymptotic region far outside the static patch, where it is ``naturally" normalized to $1$.  We prefer to think about the $L \rightarrow \infty$ limit, where $r_s$ is naturally defined in terms of the asymptotic geometry, and the derivative of the entropy with respect to $r_s$ has a well defined normalization. Another point of view is to regard the mass parameter $M$ as distinguished by the fact that it is the asymptotic energy in the $L\rightarrow\infty$ limit, and the surface gravity temperature of the cosmological horizon with the usual normalization is distinguished by the SdS first law $dS_c/dM=-1/T_c$.

Let us briefly explore other possible replica tricks using the SdS geometry.  We could introduce a boundary at some radius $r_0$ and try to define 
 \begin{align}
 n \? \frac{\beta_{SdS} \sqrt{f_{SdS}(r_0)}}{\beta_{dS} \sqrt{f_{dS}(r_0)}}.
 \end{align}
  where the $\beta$'s are the inverse cosmological horizon temperatures.  This is essentially taking the ratio of the local (Tolman) temperatures of the artificial boundary at $r_0$, as one might imagine one should do if ``the quantum system was sitting at this boundary."  It is easy to see that this proposal is not sensible, because the putative modular fluctuation computed in this manner is negative  for all $0<r_0<L$.  

Similarly, motivated by the idea that ``the hologram lives on the stretched horizon," we might try to perform a replica trick around a circle a Planck distance from the horizon,
 \begin{align}
 n \?   \frac{\beta_{SdS} \sqrt{f_{SdS}(\ell_p)}}{\beta_{dS} \sqrt{f_{dS}(\ell_p)}}.
 \end{align}
   This is similar to the previous case, except we hold the stretched horizon at a fixed proper radius  as we vary the replica index. Because the local temperature close to the horizon is nearly infinite, the heat capacity is small; however, it is still negative. One finds $-S'_1 \approx-2\pi \ell_p^2$, and again the hypothesized map from $r_s$ to the replica index is inconsistent.

We are encountering the familiar fact that if we think of any of these temperatures in the conventional way, then lowering the temperature increases the entropy, so that the specific heat is negative.  The interpretation of energy in terms of constraints on a large number of degrees of freedom in a high entropy system explains this apparent paradox\footnote{Parenthetically, the same mechanism can also be applied to the negative specific heat of black holes for non-negative c.c..}. The relevant constraint fixes the expectation value of the energy. With a fixed $\langle E \rangle$ constraint, the maximum entropy density matrix is the canonical ensemble at finite temperature $T=dS/d\langle E\rangle$. Our identification of the replica index corresponds to the ansatz that the energy of the microscopic theory describing the horizon is proportional to $-M$. In this way the constraint varies the surface gravity temperature, or equivalently the replica index. Then $-S'_1$ is equivalent to $T_c dS_c/dT_c = -dM/dT_c$, and the $-M$ can be thought of as removing energy from the cosmological horizon degrees of freedom.  

We may define a generalized heat capacity which is the sum of the heat capacity of quantum fields and the contribution from gravity,
\begin{align}
C_{gen} = C_{QFT}-S'_1.
\end{align}
When the value of $a$ obtained in effective field theory matches the value of $a$ obtained from semiclassical gravity, the generalized heat capacity so defined is proportional to the generalized entropy, $C_{gen}=a S_{gen}$. Thus it shares the properties of the generalized entropy, for example, in many cases the cutoff dependence of the quantum field contribution may be absorbed into the renormalized Newton constant. This can be anticipated from the renormalization of the Euclidean gravitational effective action on smooth manifolds.

\section{Comments on Operator Algebras}
\label{crossed}

The construction of Type $II$ crossed product algebras from the Type $III_1$ algebras of local quantum field theory and their modular automorphism groups has no {\it a priori} connection to gravitational physics.   To more easily appreciate this fact, imagine that relativistic quantum field theory had been discovered long before general relativity and no one had proposed a connection between gravitation and the geometry of space-time.  Nonetheless, the Euclidean continuation of QFT would be well-known, and physical mathematicians would have appreciated the possibility and utility of putting QFTs on a variety of Riemannian manifolds.   One can put a QFT on Euclidean Schwarzschild and 
observe that the Lorentzian continuation leads to the Hartle-Hawking state on the Kruskal manifold.  Noting that observables on the right causal patch have thermal expectation values, as expected from the Euclidean periodicity, naive field theorists will interpret this as a geometric realization of the Thermo-field Double construction.  More sophisticated theorists will note that the algebra of the right asymptotic region is Type $III_1$ and has no density matrices.  The crossed product construction can be viewed as a way of justifying this purely field theoretic conundrum of an apparent geometric picture of the TFD.   It leads to a way of calculating thermal traces, but not the total entropy, because the trace in Type $II$ algebras has a multiplicative ambiguity.  The same construction is applicable to QFT on any Riemannian manifold whose Lorentzian continuation has two causally disconnected diamonds that meet in a surface of co-dimension 2. The point that the modular crossed product is not intrinsically gravitational, and that the crossed product should be interpreted as a generic tool for regulating the entanglement entropy in arbitrary QFTs, was emphasized in~\cite{AliAhmad:2023etg,Klinger:2023tgi}.

Indeed, when one tries to calculate the entropy in any of these situations, one finds a divergence~\cite{sorkinetal} coming from short wavelength states with very small eigenvalues of the would-be modular Hamiltonian of the Type $III$ algebra.  Going back to the Riemannian calculation,  this divergence can be canceled by local counterterms involving the background geometry.  The most important of these~\cite{sussugjacob} is proportional to $\int\sqrt{g} R$ and associated boundary terms~\cite{York:1972sj,Gibbons:1976ue}.  If one interprets the Euclidean path integral as a free energy, with the time periodicity interpreted as a temperature, then varying the temperature to compute the entropy introduces singularities on the fixed surface of the time translation generator. One way to regulate it is to introduce a new boundary near that surface with associated GHY term.  Thus this counterterm tells us that both the entropy and the specific heat are cutoff-dependent and proportional to each other.  As reviewed above, direct Hilbert space calculations lead to the same conclusion.

In non-gravitational QFT, the crossed product construction leads to Type $II_{\infty}$ algebras.   In general~\cite{kfetal} the entropies assigned to the naive density matrix induced on a causal diamond by a global field theory state will have infinite entropy in this algebra, but regulated, finite entropy, density matrices can be defined, which allow one to make rigorous constructions of relative entropies and prove a number of folk theorems in the QFT literature.  The finite entropies have a finite ambiguity stemming from the multiplicative ambiguity of the trace in Type $II$ algebras.   Using the regulator in~\cite{kfetal} the entropy of QFT density matrices for local regions diverges like ${\rm ln}\ \epsilon$, while the modular fluctuation diverges like $\epsilon^{-2}$, so the ratio we have denoted by ``$a$" diverges.   This disagrees with all standard cutoff field theory calculations and with the CHM/Perlmutter prescription for CFT.   We are not sure whether this is generic or a feature of the particular regularization scheme introduced in\cite{kfetal}, or whether one should use the ambiguity in the entropy to add a term of order $\epsilon^{-2}$ to it.

The re-interpretation of the cross-product construction as a partial solution to the gravitational constraint equations in the bulk~\cite{wittenll,CLPW,CPW,JSS,KFLS} does not change any of these facts about the relation of cross products to renormalization of Euclidean path integrals for non-gravitational QFT in background geometries.\footnote{We also note that it is important to distinguish the role of the crossed product in systems with genuine symmetries, e.g. in the presence of boundaries, from its role in enforcing constraints~\cite{Klinger:2023auu}.}   The observations of\cite{maldacena} show that within the AdS/CFT correspondence, the TFD interpretation of the AdS black hole geometries are rigorously correct.  In that context, for large $AdS_d$ black holes we know that the modular fluctuation is proportional to the expectation value of the entropy, with $a = d - 2$.  In\cite{BZ} the authors explained, using the tensor network model of AdS, how this was consistent with $a = 1$ on scales small compared to the AdS radius.  

We conclude that  the crossed product construction applied to QFT in curved space-time eliminates the largest contribution to the entropy and on geometries that have a naive TFD interpretation, but does not regulate the modular fluctuation.   It does not lead to an automatic prescription for calculating gravitational entropies and modular fluctuations and any such prescription is put in by hand based on other assumptions.  In particular, it appears that~\cite{CLPW} made essentially the same assumptions about the density matrix of dS space as\cite{tbdS,wfdS,bfm}, except that the latter authors made the stronger assumption that the full Hilbert space has finite dimension.  The QFT and gravitational heat capacity arguments described in the current note add to the evidence\cite{VZ2,BZ,Banks:2022irh} that the maximal entropy assumption about the form of the density matrix may have been too strong.  

There is an enormous amount of evidence that black holes in Minkowski space are finite entropy meta-stable equilibria with extremely long lifetimes and universal properties.  The energy levels of these systems cannot have finer spacing than their inverse lifetime, which is power law in $1/S$. Yet in order to account for the entropy, the level spacing of the modular Hamiltonian must be of order $e^{-S}\ll 1/S$.   The specific heat is obviously not the modular fluctuation in this case, but it is reasonable to conjecture that there is some universal formula for it. The Carlip-Solodukhin ansatz provides such a universal formula.

Thus, within the crossed product construction, as in the earlier conjecture~\cite{tbdS,wfdS} of a finite dimensional Hilbert space for quantum theory in dS space, one must search for other principles to determine the density matrix of empty dS space.   We have shown in~\cite{Banks:2022irh} that one can retain the idea of a purely entropic definition of static energy for objects near the geodesic with essentially any finite value for the modular fluctuation coefficient $a$.   The Carlip-Solodukhin ansatz $a = 1$ has a clear but conjectural derivation from the Einstein equations near a generic causal diamond boundary, and we have provided a derivation of it from a version of the replica trick in this paper.  Quantum field theory in curved space-time gives regulator dependent answers for $a$.  All are independent of the state of the QFT.  An elegant form of regulator for CFTs gives model dependent answers, which all converge~\cite{DeBoer:2018kvc} to $a = 1$ for models with Einstein-Hilbert AdS duals, in agreement with~\cite{VZ2}.  Pauli-Villars regulators for free fields of low spin always reduce to $1 + 1$ dimensional CFTs near the boundary, again giving $a = 1$.  No field theory calculation gives $a = 0$.

We conclude these comments with an observation about the crossed product as an approximation to quantum gravity in finite space-time regions.

Much of the recent work on crossed product constructions began~\cite{wittenll} as a method of finding systematic $1/N$ corrections to the $N = \infty$ description of ``finite regions of AdS space-time" in terms of a Type $III_1$ algebra of local quantum fields\cite{LL}.  As such, one would expect the crossed product to incorporate some of the well known intuitive restrictions that gravity imposes on local quantum field theory in finite regions.  There is, as far as we can see, a fundamental problem with that idea.   The standard description of the crossed product algebra is that it consists of all linear combinations of elements of the form
\begin{equation} f_i (q) e^{ip H} a_i e^{- i p H} , \end{equation} where $H$ is the generator of the outer automorphism group of the Type $III$ algebra, $a_i$ is an element of the algebra and $f_i (q)$ is a bounded function that maps square integrable functions of $q$ into themselves.   $p$ is the canonical conjugate of $q$.  It is clear that the Type $III$ algebra is a proper sub-space of the crossed product algebra.  The Type $II_1$ algebra that has been suggested for dS space simply projects this algebra onto the positive spectrum of $q$ and still contains the QFT algebra as a proper sub-space.  It is not a sub-algebra because the projection operator distorts the operator product.  

It has long been argued that the gravitational backreaction produced by the expectation value of the stress tensor 
\begin{equation} \langle 0 | a^{\dagger} T_{mn} (x) a | 0 \rangle , \end{equation} for most of the operators $a$ in the algebra of a finite region in Minkowski space, would distort that geometry and create large black holes.  Such states, however, can be  excised from the theory without causing any conflict with experiment~\cite{Cohen:1998zx,Banks:2019arz,Blinov:2021fzl}. This suggests that the correct quantum gravity algebra of a finite space time region is smaller, rather than larger, than the Type $III$ algebra assigned to the region by QFT.

\section{Discussion}
\label{disc}
We have argued that a reasonable interpretation of semiclassical evidence is that 
empty de Sitter space is described by a finite dimensional quantum theory in a state with a large but not maximal entropy. Deviations from maximal uncertainty are associated with a nonzero modular fluctuation, or a heat capacity in thermal states, which in cutoff quantum field theory is of order the entropy. In semiclassical gravity a natural conjecture for the relevant heat capacity provides the same result.

Of course, this is only an observation, not a proof. Even if each term in the generalized heat capacity accurately computes an energy fluctuation in the corresponding subsystem, the sum may not share this interpretation for the whole system. An elementary example is an isolated box of gas in the microcanonical state, divided into two subsystems by a heat-conducting membrane.

The apparent tension with crossed product proposals is not fully resolved. Of course, it may be that they are simply computing  approximations to different quantities.

Let us conclude by describing connections to related ideas in holography. The Covariant Entropy Bound (CEB) for causal diamonds suggests a conjecture that physics inside diamonds can be described by models with finite dimensional Hilbert spaces obeying an area law~\cite{raphreview}. A ``natural" state is the maximal entropy density matrix on that Hilbert space, and other guiding principles are not immediately apparent in the absence of space-time isometries. Indeed, various prior works by Banks and Fischler viewed the maximal entropy conjecture for de Sitter space as a special case of this reasoning for general causal diamonds. 
  However, such a principle had already been found~\cite{Carlip,solo} although its validity for general causal diamonds was not recognized until much later~\cite{BZ}.  According to this principle, the modular Hamiltonian of a causal diamond is the Virasoro generator $L_0$ of a cut-off $1 + 1$ dimensional CFT living on an interval.  The CFT has a central charge proportional to the area of the diamond's holographic screen, and Cardy's formula reproduces the Bekenstein-Hawking area law for entropy of black holes and causal diamonds. Ref.~\cite{BZ} interpreted~\cite{Carlip,solo} as a principle about the microscopic theory and argued that it implies $a=1$. The present work establishes a level of semiclassical compatibility.

The Carlip-Solodukhin ansatz also arose in an entirely different context\cite{mannelli}, in an attempt to model the dynamics of an FRW cosmology that saturated the CEB at all times.  This is a flat FRW model with equation of state that is a mixture of $p = \pm \rho$.  It asymptotes to dS space and predicts that the dS modular Hamiltonian is not a c-number.  In this context, the fact that the modular fluctuations are proportional to the entropy is important for explaining the Cosmic Microwave Background fluctuations\cite{holoinflationrevised}\cite{tbas}.

An additional choice needs to be made in order to extend the semiclassical heat capacity of de Sitter space studied here to smaller causal diamonds. This is because the first law of causal diamonds contains a term proportional to $k dV$~\cite{Jacobson:2018ahi}, where $V$ is a spatial volume and $k$ is a codimension-2 extrinsic curvature trace. $k$ vanishes for maximal diamonds like the static patch of de Sitter, but not for more general diamonds. When we compute $TdS/dT$ for de Sitter, we are effectively defining the diamond for different $T$ by fixing the c.c. and fixing $k=0$. Thus a natural extension to non-maximal diamonds seems to be holding $k$ (the Brown-York surface energy density) fixed but nonzero. This question will be explored elsewhere.

~\\

{\bf Acknowledgments }\\
We thank Shadi Ali Ahmad, Ro Jefferson, Marc Klinger, Rob Leigh, Horacio Casini, and Antony Speranza for discussions and comments on the manuscript. The work of TB  is partially supported by the Department of Energy under grant DOE SC0010008. PD acknowledges support from the U.S. 
Department of Energy, Office of Science, Office of High Energy Physics under award number DE-SC0015655.

~\\

\end{document}